\documentclass[journal]{IEEEtran}

\usepackage{cite}
\usepackage{graphicx}
\usepackage{hyperref}
\usepackage{amsmath}
\usepackage{amssymb}
\usepackage{array}
\usepackage{url}
\usepackage{color}

\begin{document}
\title{Sidelobe Modification for Reflector Antennas by Electronically-Reconfigurable Rim Scattering}
\author{
S.W.~Ellingson,~\IEEEmembership{Senior Member,~IEEE}
and
R.~Sengupta
\thanks{The authors are with the Dept.\ of Electrical and Computer Engineering, Virginia Tech, Blacksburg, VA, 24061 USA, e-mail: ellingson.1@vt.edu.}
\thanks{This material is based upon work supported in part by the National Science Foundation under Grant ECCS 2029948.}
\thanks{\copyright 2021 IEEE.  Personal use of this material is permitted.  Permission from IEEE must be obtained for all other uses, in any current or future media, including reprinting/republishing this material for advertising or promotional purposes, creating new collective works, for resale or redistribution to servers or lists, or reuse of any copyrighted component of this work in other works.}
\thanks{Manuscript accepted for publication in IEEE Antennas \& Wireless Propagation Letters.}}
%
\maketitle

\begin{abstract}
Dynamic modification of the pattern of a reflector antenna system traditionally requires an array of feeds.  This paper presents an alternative approach in which the scattering from a fraction of the reflector around the rim is passively modified using, for example, an electronically-reconfigurable reflectarray.  This facilitates flexible sidelobe modification, including sidelobe canceling, for systems employing a single feed.  
Applications for such a system include radio astronomy, where deleterious levels of interference from satellites enter through sidelobes.  
We show that an efficient reconfigurable surface occupying about 11\% of the area 
of an axisymmetric circular paraboloidal reflector antenna fed from the prime focus is sufficient to null interference arriving from any direction outside the main lobe with 
little change in the main lobe characteristics.  
We further show that the required surface area is independent of frequency and that the same performance can be obtained using 1-bit phase control of the constituent unit cells for a reconfigurable surface occupying an additional 6\% of the reflector surface.    
\end{abstract}


%

\section{Introduction}
\label{sIntro}

\IEEEPARstart{R}{eflector}  
antenna systems are commonly used when high gain is required to receive weak signals.  In these applications, the system is vulnerable to interference arriving through sidelobes.   
This vulnerability can be ameliorated by sidelobe modification; and in particular, by sidelobe canceling. 
Sidelobe canceling involves placing a pattern null within the sidelobe so as to reject the interference, and ideally without significantly affecting the main lobe.  
Traditional sidelobe canceling requires an array \cite{MM80}.  
Such an array can be implemented in a reflector antenna system using an array of feeds in lieu of the traditional single feed; see e.g.\ \cite{Bird16} 
and references therein.  
However feed arrays entail compromises including 
greater aperture blockage and
dynamic variability in gain and the shape of the main lobe.      

An example is radio astronomy, where telescopes are commonly implemented as either large reflectors 
or arrays of large reflectors (see e.g., \cite{E16}). 
These systems are vulnerable to interference from satellite transmissions received through sidelobes.  
Of particular concern are transmissions at frequencies near astrophysical spectral lines near 1.6~GHz and 10.7~GHz. 
This long-standing problem is expected to become worse in coming years as new generations of satellites begin transmitting at these frequencies; see e.g., \cite{CEPT-ECC-271, UNOOSA-IAU-2021}. 
Most radio telescopes now operational or planned do not employ feed arrays, as such arrays are not scientifically necessary in most applications and introduce onerous performance limitations in applications where they are not required.  
In particular, top-tier radio telescopes such as the Expanded Very Large Array (VLA) \cite{EVLA} and Green Bank Telescope (GBT) \cite{GBT}, and new telescopes in development including Next Generation VLA (ngVLA) \cite{ngVLA} and Deep Synoptic Array (DSA) \cite{DSA} use single-feed systems.

\begin{figure}
\centerline{\includegraphics[width=0.8\columnwidth]{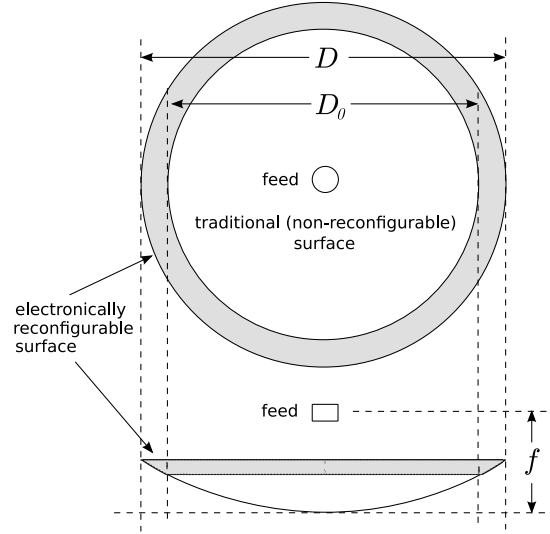}}
\caption{
On-axis (\emph{top}) and side (\emph{bottom}) views of an electronically-reconfigurable rim scattering system. 
\label{fConcept}
}
\end{figure}

This paper presents an alternative approach to sidelobe modification which can be implemented in single-feed systems.  
The concept is illustrated in Figure~\ref{fConcept}.
For simplicity, we limit the scope of this paper to prime focus-fed circular axisymmetric paraboloidal reflectors; however the concept is applicable to other types of systems, including those employing Cassegrain or Gregorian optics, and those using shaped reflectors. 
In this concept, scattering from a fraction of the reflector surface around the rim is electronically modified by manipulating the phase of scattering from unit cells comprising the surface.
The unit cells are envisioned to be contiguous elements having sub-wavelength dimension.
For example, the surface could be implemented as a reflectarray, with unit cells implemented as patch antennas whose scattering is controlled by manipulating the impedances presented to the antenna terminals \cite{Hum+1_1401}.
 
The concept of pattern modification by altering the reflector surface is not entirely new.  
``Rim loading'' with a fixed surface impedance as a means to reduce sidelobe levels is described in  
\cite{BDS81}.
The use of fixed mechanical modification of the surface as a means to optimize sidelobes is described in \cite{CH17},
and as a means to implement dual-frequency beam optimization is described in \cite{C+20}.
Novel aspects of the present work include
(1) electronic reconfigurability, enabling dynamic pattern modification; 
(2) analysis addressing area that must be given over to reconfigurability; and
(3) analysis demonstrating that the method is robust to errors in control and hardware failure.

\section{Method of Analysis}
\label{sMOA}

\begin{figure}
\centerline{\includegraphics[width=0.8\columnwidth]{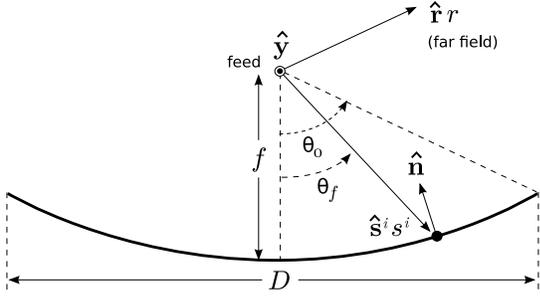}}
\caption{
Geometry for analysis of reflector antenna systems in this paper.  
\label{fDishGeo}
}
\end{figure}

The directivity, main lobe shape, and characteristics of the largest sidelobes of an electrically-large reflector antenna can be accurately determined using the theory of physical optics (PO; see e.g.\ \cite{ST13}).  
Consider the case shown in Figure~\ref{fDishGeo} of an axisymmetric paraboloidal reflector with a single feed located at its focus.  
The receive directivity and pattern of this system is equal to the transmit directivity and pattern, and the transmit pattern is determined using the following procedure:

\begin{enumerate}

\item Calculate the field incident at the surface (${\bf s}^i = \hat{\bf s}^i s^i$) of the reflector, due to the feed. 

\item Calculate the PO equivalent surface current distribution
${\bf J}_0 = 2\hat{\bf n}\times{\bf H}^i$
where 
$\hat{\bf n}$ is the unit normal to the surface and 
${\bf H}^i$ is the incident magnetic field intensity.

\item In the far field, the electric field intensity ${\bf E}^s$ scattered by the reflector is determined by integrating over the reflector:
\begin{equation}
{\bf E}^s = -j\omega\mu_0 \frac{e^{-jkr}}{4\pi r} 
            \int_{\theta_f=0}^{\theta_0} \int_{\phi=0}^{2\pi}  
            {\bf J}_0({\bf s}^i) 
            ~e^{-jk\hat{\bf r}\cdot{\bf s}^i} 
            ds 
\label{ePO}
\end{equation}
where 
$j=\sqrt{-1}$, 
$\omega$ is $2\pi$ times frequency, 
$\mu_0$ is the permeability of free space, 
$k$ is wavenumber, 
$\hat{\bf r}$ points from the origin of the global coordinate system toward the field point,
$\theta_f$ is the angle measured from the reflector axis of rotation toward the rim (thus, $\theta_f=\theta_0$ at the rim),
$\phi$ is the angular coordinate orthogonal to both $\theta_f$ and the reflector axis, 
and
$ds$ is the differential element of surface area. 

\item The far-field power pattern $P(\hat{\bf r})$ is determined in the traditional way by eliminating the radial ($\hat{\bf r}$) component of ${\bf E}^s$, computing the resulting power density, and dividing by the power density averaged over a sphere bounding the system.  This yields units of directivity.

\end{enumerate}

To establish a baseline of performance, consider a traditional reflector antenna system having
diameter $D=18$~m and
focal ratio $f/D=0.4$,
with a feed modeled 
as follows:
\begin{equation}
{\bf H}^i({\bf s}^i) = I_0 \frac{        \hat{\bf y} \times \hat{\bf s}^i }
                     { \left| \hat{\bf y} \times \hat{\bf s}^i \right| } 
                \frac{ e^{-jks^i} }
                     {       s^i  } \left( \cos{\theta_f} \right)^q ~\mbox{,~~~}\theta_f \le \pi/2
\label{eFeed}                     
\end{equation} 
and ${\bf H}^i=0$ for $\theta_f > \pi/2$.
In this model, $I_0$ represents the source magnitude and phase and $q$ is used to set the directivity of the feed. 
Setting $q=1.14$ yields edge illumination (EI; i.e., ratio of field intensity in the direction of the rim to the field intensity in the direction of the vertex), to approximately $-11$~dB, yielding aperture efficiency of about 81.5\%.  

\begin{figure}
\centerline{\includegraphics[width=\columnwidth]{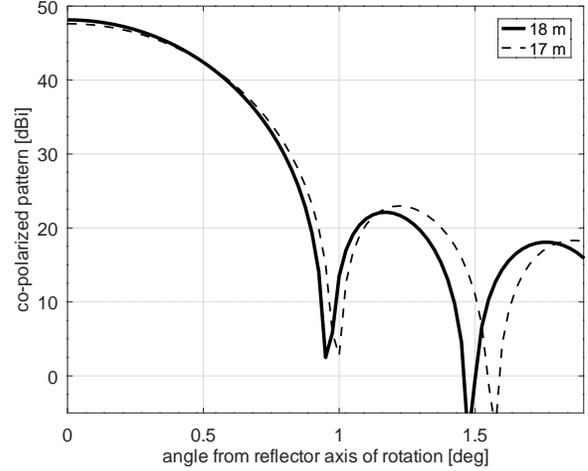}}
\caption{
\emph{Solid:} H-plane co-polarized pattern of the reflector system described in Section~\ref{sMOA}, $D=18$~m, 1.5~GHz.
\emph{Dashed:} Same, except diameter reduced to $17$~m (addressed in Section~\ref{sFSC}).
\label{fEx1}
}
\end{figure}

Figure~\ref{fEx1} shows the H-plane pattern of this system at 1.5~GHz, limiting view to the first few sidelobes around the main lobe where the PO approximation is known to be accurate.  
The PO integral was computed numerically using approximately square surface elements having side length $0.2\lambda$, 
where $\lambda$ is wavelength. 
Directivity and first sidelobe level are found to be $48.2$~dBi and $-26$~dB, respectively.

Next we consider a system which is identical except that a reconfigurable surface replaces the portion of the reflector surface between radial distances 8.5~m and 9~m.
The new system may be viewed as a traditional reflector having diameter $D_0=17$~m, with a reconfigurable surface that follows the same paraboloidal surface and extends the diameter to $D=18$~m.
In this case the scattered field is
${\bf E}^s = {\bf E}^s_0 + {\bf E}^s_1$
where 
${\bf E}^s_0$ is the field scattered by the non-reconfigurable center portion of the reflector, and 
${\bf E}^s_1$ is the field scattered by the reconfigurable surface.
The former is given by Equation~\ref{ePO} with $\theta_0=\theta_1$, where $\theta_1$ is the angle to the rim of the non-reconfigurable part of the reflector.
Similarly, the field scattered by the reconfigurable surface can be calculated as follows:
\begin{equation}
{\bf E}^s_1 = -j\omega\mu_0 \frac{e^{-jkr}}{4\pi r} 
            \int_{\theta_f=\theta_1}^{\theta_0} \int_{\phi=0}^{2\pi}  
            {\bf J}_1({\bf s}^i) 
            ~e^{-jk\hat{\bf r}\cdot{\bf s}^i} 
            ds 
\label{ePOe}
\end{equation}
where ${\bf J}_1$ is the equivalent surface current for scattering from the reconfigurable surface.

The function ${\bf J}_1$ depends on the technology used to modify the scattering and the state of the reconfigurable surface.  
As an initial assessment of feasibility, it is useful to be able to calculate estimates of performance that are independent of the technology used to implement the surface.  
To accomplish this, the reconfigurable surface is modeled as a contiguous surface of approximately square flat plates having side length $0.5\lambda$ (area $\Delta s=0.25\lambda^2$). 
These plates represent the unit cells.
Scattering is calculated using Equation~\ref{ePOe}, except now quantizing the integrand to the dimensions of the plates.
Thus:
\begin{equation}
{\bf E}^s_1 = -j\omega\mu_0 \frac{e^{-jkr}}{4\pi r} 
            \sum_n  
            {\bf J}_1({\bf s}^i_n) 
            ~e^{-jk\hat{\bf r}\cdot{\bf s}^i_n} 
            \Delta s 
\label{ePOed}
\end{equation}
where ${\bf s}^i_n$ is the center of cell $n$, and $n$ indexes the cells.
Further, we assume
\begin{equation}
{\bf J}_1({\bf s}^i_n) = c_n {\bf J}_0({\bf s}^i_n)
\label{eRSJ}
\end{equation}
that is, ${\bf J}_1$ is the PO surface current at the center of the plate, times a complex constant $c_n$ to account for reconfigurability (e.g., phase shifting), imperfect efficiency, and any other effects associated with whatever technology is used to implement the surface.  
This is similar to the physics-based methods used in \cite{E19} and \cite{OBL19} to model reconfigurable scattering surfaces in other applications.
The motivation for $0.5\lambda$ quantization of the surface is simply that technologies that might be used to implement reconfigurability would normally consist of unit cells having approximately this periodicity in order to satisfy the Nyquist condition for full sampling of the available aperture.

In a practical system, there should be negligible modification of the main lobe when the reconfigurable surface is uncontrolled; in particular, when $c_n = 1$ for all cells, either by choice or due to system failure.  In this case the only difference between the original ($D_0=D=18$~m) system and the modified ($D=18$~m, $D_0=17$~m) system is that the outer 0.5~m of the reflector surface consists of 2756 contiguous half-wavelength-square flat plates (representing the uncontrolled cells) as opposed to a continuously-varying surface.  The result in this case is not significantly different from the result shown in Figure~\ref{fEx1}, and can be detected only as a $\sim 0.1$~dB difference in the peak level of the second sidelobe.

\section{Feasibility for Sidelobe Modification}
\label{sFSC}

This section addresses the following question:
How much of the surface of the reflector must be given over to reconfigurability in order to obtain effective sidelobe modification?
Consider the patterns $P_0(\hat{\bf r})$ and $P_1(\hat{\bf r})$ associated with ${\bf E}^s_0$ and ${\bf E}^s_1$, respectively.
For full ability to modify sidelobes over the entire angular span outside the main lobe, it must be possible to achieve a value of $P_1$ which is at least as large as $P_0$ over this span.
Let $P_{1m}(\hat{\bf r})$ be the largest value of $P_1(\hat{\bf r})$ that can be achieved by manipulating the reconfigurable surface.
Then, canceling of interference located at the peak of the largest sidelobe requires the $P_{1m}(\hat{\bf r})$ to be at least as large as $P_0(\hat{\bf r})$ in the direction of this peak.
If this condition is satisfied, then the area allocated to the reconfigurable surface is sufficient in the sense that any technology and control algorithm providing sufficiently fine control of the scattered field can be used to implement effective sidelobe canceling.  

Thus, $P_{1m}(\hat{\bf r})$ is of primary interest.  
Note that $P_{1m}$ is not a pattern corresponding to a single set of $c_n$'s (i.e., $P_1$), but rather the value of $P_1$ for whatever choice of $c_n$'s maximize $P_1$ in each direction.
Since $P_1(\hat{\bf r})$ varies monotonically with the magnitude of the field scattered by the reconfigurable surface, $P_{1m}(\hat{\bf r})$ is obtained when all contributions to the associated field integral add in phase in direction $\hat{\bf r}$.
This can be determined from the scattered field calculated from Equations~\ref{ePOed} and \ref{eRSJ} with
\begin{equation}
c_n = \exp\left\{ -j \arg\left[ {\bf J}_0({\bf s}^i_n) ~e^{-jk\hat{\bf r}\cdot{\bf s}^i_n} \right] \right\}
\label{ecnmax}
\end{equation}
In other words, the field scattered from each cell is assumed to be phase shifted such that all terms of the sum add in phase.  
This assumes that the phase of each reconfigurable element is continuously variable; the effect of phase quantization will be addressed in Section~\ref{sPQ}.
This also assumes that the reconfigurable surface is 100\% efficient; i.e., all incident power is scattered.
Thus we obtain an upper limit on performance which depends only on the area given over to reconfigurability.

\begin{figure}
\centerline{\includegraphics[width=\columnwidth]{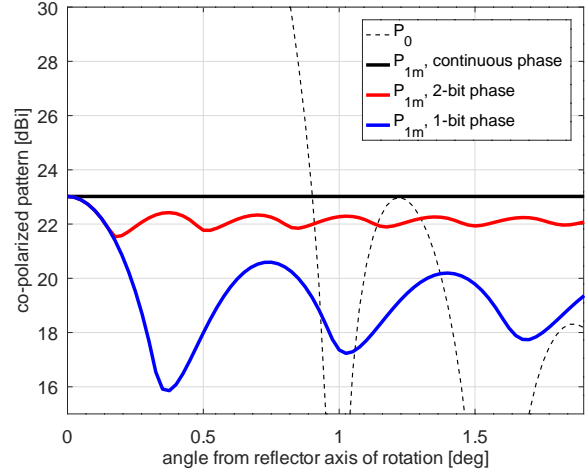}}
\caption{
Pattern $P_0$ for the $D_0=17$~m center region and maximum pattern $P_{1m}$ for the 0.5~m-wide reconfigurable region around the rim. 
\label{fEx2}
}
\end{figure}

To demonstrate, let us again consider the $D=18$~m, $D_0=17$~m example, but now computing $P_0$ and $P_{1m}$ separately, with $P_{1m}$ determined from scattered fields calculated via Equations~\ref{ePOed}--\ref{ecnmax}.
First, consider the dashed (17~m) line in Figure~\ref{fEx1}, which is $P_0$ in this case.  It is the sidelobes of this pattern that the reconfigurable surface is attempting to modify.
The performance of new system is shown in Figure~\ref{fEx2}.
Here we observe that the continuously-variable phase shifting scheme represented by Equation~\ref{ecnmax} yields $P_{1m} = P_0$ at the peak of the first sidelobe.
Thus, 
10.8\% of the surface area of the original $D_0=18$~m reflector must be given over to reconfigurability to meet the $P_{1m}\ge P_0$ criterion in this case.

This demonstration reveals an attractive feature of the proposed method:  
The \emph{inability} to significantly change the main lobe.  
From Figures~\ref{fEx1} and \ref{fEx2} we see that $P_{1m}$ is about $-26$~dB (0.3\%) relative to $P_0$ at the center of the main lobe.  
This upper-bounds the degradation and variability that the reconfigurable surface is able to impart on the main lobe, regardless of the state (i.e., $c_n$'s) of the reconfigurable surface. Since the reconfigurable surface is passive, this is true regardless of any failures of individual cells; i.e., the scattering from malfunctioning cells cannot be significantly greater than that from functioning cells.  
Therefore, control error and system failure cannot significantly alter the directivity and shape of the main lobe.
Further, the very large number of degrees of freedom in this system allow many constraints to be enforced on $P_1$. 
For example, the system could null $P_1$ along the reflector axis 
while simultaneously imposing other constraints on $P_1$ in order to accomplish the desired sidelobe modifications.

\section{Phase Quantization}
\label{sPQ}

Practical reconfigurable surfaces may require quantization of phase control.
That is, it may be necessary to limit the phase of $c_n$ to discrete values.
Figure~\ref{fEx2} shows the effect of phase quantization.
For 1-bit phase quantization, $c_n$ is either $+1$ or $-1$, whichever results in the phase of the associated term being closest to $0$.
For 2-bit phase quantization, $c_n$ is either $+1$, $+j$, $-1$, or $-j$. 
As expected, phase quantization generally reduces $P_{1m}$.  

To offset the degradation associated with phase quantization, it is necessary to increase the area of the reflector given over to reconfigurability.
In the example system, it is found that $D_0=16.4$~m (about 17\% of the surface made reconfigurable) is needed for 1-bit phase quantization to yield $P_{1m}=P_0$ at the peak of the first sidelobe of $P_0$.
Since the number of reconfigurable unit cells is large, 
the quantization of unit cell phase is not expected to significantly limit the accuracy to which the phase of the field scattered from the reconfigurable surface can be set. 

\begin{figure}
\centerline{\includegraphics[width=\columnwidth]{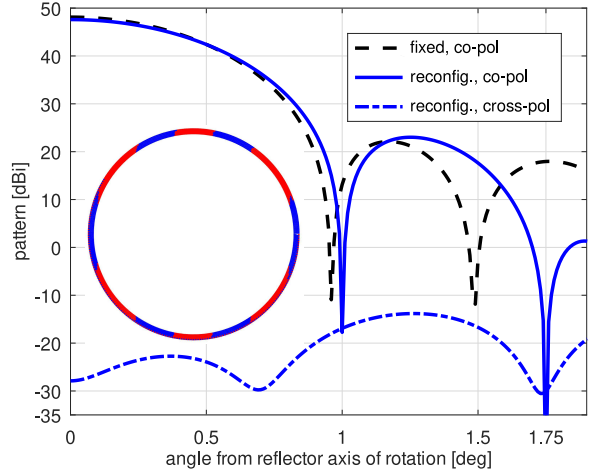}}
\caption{
Total patterns for the $D=D_0=18$~m (non-reconfigurable) system and the $D=18$~m, $D_0=17$~m reconfigurable system with 1-bit phase quantization.
The inset is an on-axis view of the system with the surface of the reconfigurable rim color-coded to indicate unit cell state. 
\label{fEx5}
}
\end{figure}

Figure~\ref{fEx5} shows an example of the performance of the $D_0=17$~m system using 1-bit quantization.
The system is configured to create a null at $1.75^{\circ}$ in the co-polarized pattern, which is close to the peak of the second sidelobe of the unmodified ($D_0=18$~m) system.
Figure~\ref{fEx2} indicates that a null at this location is within the capability of the system.
For this example, the following simple ``blind'' algorithm is used to determine the $c_n$'s:  
We compute the pattern in the direction of the desired null by summing the contributions from unit cells one at a time, but select whichever value of $c_n$ ($+1$ or $-1$) minimizes the magnitude of the accumulated co-polarized field at that step in the integration.
These $c_n$'s are then used to compute the entire pattern.
Figure~\ref{fEx5} confirms that this method does indeed yield a deep null.

The $c_n$'s for this example are shown in the inset of Figure~\ref{fEx5}.
Note that the distribution of $c_n$'s is highly structured and largely contiguous, suggesting the existence of methods for deterministic initialization and rapid updating of these values. 

Also shown in Figure~\ref{fEx5} is the resulting cross-polarized pattern.
Since cross-polarization is ideally zero in this plane, we see that the unconstrained optimization of the reconfigurable surface has degraded cross-polarization performance.

\section{Edge Illumination}
\label{sEI}

For the system considered in previous sections, EI $=-11$~dB, which is known to optimize aperture efficiency for prime focus-fed axisymmetric paraboloidal reflector systems with feeds of the type described by Equation~\ref{eFeed} \cite{ST13}.
In some applications, it is desirable to further reduce EI, which reduces sidelobe levels at the expense of aperture efficiency.
A system with reduced EI will exhibit generally lower $P_{1m}$, since less power will be incident on the rim.  
Thus, it is of interest to consider the efficacy of reconfigurable rim surface scattering for systems with lower EI.

To this end, we repeated the analysis of the reflector addressed in Section~\ref{sFSC} with the feed parameter $q$ increased to 1.85, which decreases EI to $-16$~dB.
In this case we find that $P_{1m}$ is reduced from about 23~dB to about 20~dB for continuously-variable phase unit cells. 
However $P_0$ is also reduced, and in fact the margin $P_{1m}/P_0$ for the peak of the first sidelobe is \emph{increased} from about 0~dB to about 2~dB.
Therefore the proposed method does not require any particular level of edge illumination for effective operation.

\section{Frequency Considerations}
\label{sFreqCon}

Results shown in previous sections were generated for 1.5~GHz.
For reasons addressed in Section~\ref{sIntro}, higher frequencies are also of interest.
The principal difference for implementation at higher frequencies is that unit cells of the reconfigurable surface, assumed to be $0.5\lambda$ square, will be correspondingly smaller.  
However, the peak sidelobe level of $P_0$ depends only on feed pattern (e.g., EI) and is otherwise independent of frequency.
Therefore, the total area required for the reconfigurable surface will be independent of frequency as long as the feed pattern is independent of frequency.
Subsequently, the required number of unit cells comprising the surface increases in proportion to frequency squared.

A separate question is whether a reconfigurable surface comprised of unit cells optimized for a higher frequency, such as 10.7~GHz, would also be effective at a lower frequency, such as 1.5~GHz.  This depends on the technology as well as the specific design of the reconfigurable surface, and is left as a topic for future investigation.   

\section*{Acknowledgment}

The authors acknowledge R.M.\ Buehrer of Virginia Tech for suggesting the algorithm described in Section~\ref{sPQ}.



\end{document}